\PassOptionsToPackage{unicode}{hyperref}
\PassOptionsToPackage{hyphens}{url}
\PassOptionsToPackage{dvipsnames,svgnames,x11names}{xcolor}
\documentclass[
]{article}
\usepackage{xcolor}
\usepackage[margin=1in]{geometry}
\usepackage{amsmath,amssymb}
\usepackage{iftex}
\ifPDFTeX
  \usepackage[T1]{fontenc}
  \usepackage[utf8]{inputenc}
  \usepackage{textcomp} 
\else 
  \usepackage{unicode-math} 
  \defaultfontfeatures{Scale=MatchLowercase}
  \defaultfontfeatures[\rmfamily]{Ligatures=TeX,Scale=1}
\fi
\usepackage{lmodern}
\ifPDFTeX\else
\fi
\IfFileExists{upquote.sty}{\usepackage{upquote}}{}
\IfFileExists{microtype.sty}{
  \usepackage[]{microtype}
  \UseMicrotypeSet[protrusion]{basicmath} 
}{}
\makeatletter
\@ifundefined{KOMAClassName}{
  \IfFileExists{parskip.sty}{%
    \usepackage{parskip}
  }{
    \setlength{\parindent}{0pt}
    \setlength{\parskip}{6pt plus 2pt minus 1pt}}
}{
  \KOMAoptions{parskip=half}}
\makeatother
\makeatletter
\ifx\paragraph\undefined\else
  \let\oldparagraph\paragraph
  \renewcommand{\paragraph}{
    \@ifstar
      \xxxParagraphStar
      \xxxParagraphNoStar
  }
  \newcommand{\xxxParagraphStar}[1]{\oldparagraph*{#1}\mbox{}}
  \newcommand{\xxxParagraphNoStar}[1]{\oldparagraph{#1}\mbox{}}
\fi
\ifx\subparagraph\undefined\else
  \let\oldsubparagraph\subparagraph
  \renewcommand{\subparagraph}{
    \@ifstar
      \xxxSubParagraphStar
      \xxxSubParagraphNoStar
  }
  \newcommand{\xxxSubParagraphStar}[1]{\oldsubparagraph*{#1}\mbox{}}
  \newcommand{\xxxSubParagraphNoStar}[1]{\oldsubparagraph{#1}\mbox{}}
\fi
\makeatother

\usepackage{longtable,booktabs,array}
\usepackage{calc} 
\usepackage{etoolbox}
\makeatletter
\patchcmd\longtable{\par}{\if@noskipsec\mbox{}\fi\par}{}{}
\makeatother
\IfFileExists{footnotehyper.sty}{\usepackage{footnotehyper}}{\usepackage{footnote}}
\makesavenoteenv{longtable}
\usepackage{graphicx}
\makeatletter
\newsavebox\pandoc@box
\newcommand*\pandocbounded[1]{
  \sbox\pandoc@box{#1}%
  \Gscale@div\@tempa{\textheight}{\dimexpr\ht\pandoc@box+\dp\pandoc@box\relax}%
  \Gscale@div\@tempb{\linewidth}{\wd\pandoc@box}%
  \ifdim\@tempb\p@<\@tempa\p@\let\@tempa\@tempb\fi
  \ifdim\@tempa\p@<\p@\scalebox{\@tempa}{\usebox\pandoc@box}%
  \else\usebox{\pandoc@box}%
  \fi%
}
\def\fps@figure{htbp}
\makeatother

\NewDocumentCommand\citeproctext{}{}
\NewDocumentCommand\citeproc{mm}{%
  \begingroup\def\citeproctext{#2}\cite{#1}\endgroup}
\makeatletter
 \let\@cite@ofmt\@firstofone
 \def\@biblabel#1{}
 \def\@cite#1#2{{#1\if@tempswa , #2\fi}}
\makeatother
\newlength{\cslhangindent}
\newlength{\csllabelwidth}
\newenvironment{CSLReferences}[2] 
 {\begin{list}{}{%
  \setlength{\itemindent}{0pt}
  \setlength{\leftmargin}{0pt}
  \setlength{\parsep}{0pt}
  \ifodd #1
   \setlength{\leftmargin}{\cslhangindent}
   \setlength{\itemindent}{-1\cslhangindent}
  \fi
  \setlength{\itemsep}{#2\baselineskip}}}
 {\end{list}}
\usepackage{calc}

\usepackage{booktabs}
\usepackage{longtable}
\usepackage{array}
\usepackage{multirow}
\usepackage{wrapfig}
\usepackage{float}
\usepackage{colortbl}
\usepackage{pdflscape}
\usepackage{tabu}
\usepackage{threeparttable}
\usepackage{threeparttablex}
\usepackage[normalem]{ulem}
\usepackage{makecell}
\usepackage{xcolor}
\usepackage{orcidlink}
\makeatletter
\@ifpackageloaded{caption}{}{\usepackage{caption}}
\AtBeginDocument{%
\ifdefined\contentsname
  \renewcommand*\contentsname{Table of contents}
\else
  \newcommand\contentsname{Table of contents}
\fi
\ifdefined\listfigurename
  \renewcommand*\listfigurename{List of Figures}
\else
  \newcommand\listfigurename{List of Figures}
\fi
\ifdefined\listtablename
  \renewcommand*\listtablename{List of Tables}
\else
  \newcommand\listtablename{List of Tables}
\fi
\ifdefined\figurename
  \renewcommand*\figurename{Figure}
\else
  \newcommand\figurename{Figure}
\fi
\ifdefined\tablename
  \renewcommand*\tablename{Table}
\else
  \newcommand\tablename{Table}
\fi
}
\@ifpackageloaded{float}{}{\usepackage{float}}
\floatstyle{ruled}
\@ifundefined{c@chapter}{\newfloat{codelisting}{h}{lop}}{\newfloat{codelisting}{h}{lop}[chapter]}
\floatname{codelisting}{Listing}

\makeatother
\makeatletter
\@ifpackageloaded{caption}{}{\usepackage{caption}}
\@ifpackageloaded{subcaption}{}{\usepackage{subcaption}}
\makeatother
\usepackage{bookmark}
\IfFileExists{xurl.sty}{\usepackage{xurl}}{} 
\hypersetup{
  pdftitle={Covariate-Balanced Weighted Stacked Difference-in-Differences},
  pdfauthor={Vadim Ustyuzhanin},
  pdfkeywords={difference-in-differences, stacked
DID, matching, weighting, staggered adoption, panel data},
  colorlinks=true,
  linkcolor={blue},
  filecolor={Maroon},
  citecolor={blue},
  urlcolor={Blue},
  pdfcreator={LaTeX via pandoc}}

\title{Covariate-Balanced Weighted Stacked Difference-in-Differences}

\author{
\textbf{Vadim Ustyuzhanin}~\orcidlink{0000-0003-3800-1108}\\
HSE University\\
\href{mailto:vvustiuzhanin@hse.ru}{vvustiuzhanin@hse.ru}
}

\date{April 2, 2026}

\begin{document}
\maketitle
\begin{abstract}
This paper proposes Covariate-Balanced Weighted Stacked
Difference-in-Differences (CBWSDID), a design-based extension of
weighted stacked DID for settings in which untreated trends may be
conditionally rather than unconditionally parallel. The estimator
separates within-subexperiment design adjustment from
across-subexperiment aggregation: matching or weighting improves
treated-control comparability within each stacked subexperiment, while
the corrective stacked weights of Wing et al.~recover the target
aggregate ATT. I show that the same logic extends from absorbing
treatment to repeated \(0 \to 1\) and \(1 \to 0\) episodes under a
finite-memory assumption. The paper develops the identifying framework,
discusses inference, presents simulation evidence, and illustrates the
estimator in applications based on Trounstine (2020) and Acemoglu et
al.~(2019). Across these examples, CBWSDID serves as a bridge between
weighted stacked DID and design-based panel matching. The accompanying
\textbf{R package} \texttt{cbwsdid} is available on
\href{https://github.com/vadvu/cbwsdid}{GitHub}.
\end{abstract}

\section{Introduction}\label{introduction}

Stacked difference-in-differences has become a widely used way to study
dynamic treatment effects in staggered-adoption settings. Its appeal is
clear: by reorganizing the data into cohort-specific sub-experiments, it
avoids many of the comparison problems that arise in conventional TWFE
event-study regressions. Yet stacked DID still leaves two distinct
design problems. First, the final estimator must aggregate treated and
control observations across sub-experiments in a way that targets a
well-defined causal parameter. Wing et al.
(\citeproc{ref-wing2024}{2024}) show that ordinary stacked DID fails on
this margin and propose corrective stacked weights that recover the
trimmed aggregate ATT. Second, even within a given sub-experiment,
treated and control units may differ substantially on lagged outcomes or
other pre-treatment characteristics. Correct aggregation across
sub-experiments does not by itself guarantee credible treated-control
comparisons within them.

This paper proposes Covariate-Balanced Weighted Stacked DID (CBWSDID),
which separates these two tasks. In the first stage, matching or
weighting is used to improve comparability within each stacked
sub-experiment. In the second stage, the corrective weighting logic of
Wing et al. (\citeproc{ref-wing2024}{2024}) is applied to recover the
desired aggregate ATT. The key point is that these steps can be combined
without treating matching and weighting as separate estimators. Both are
represented through nonnegative first-stage design weights, which makes
it possible to refine the donor pool within each sub-experiment while
preserving the target estimand defined by weighted stacked DID.

The paper makes three contributions. First, it provides a unified
stacked DID framework that accommodates matching-based and
weighting-based refinement within the same estimator. Second, it extends
this logic beyond absorbing treatment to repeated \(0 \to 1\) episodes
under a finite-memory assumption and, by symmetry, to \(1 \to 0\)
episodes as well. This creates a bridge between weighted stacked DID and
episode-based designs such as Imai et al.
(\citeproc{ref-imai2023}{2023}). Third, it develops the corresponding R
package \texttt{cbwsdid} (available on
\href{https://github.com/vadvu/cbwsdid}{GitHub}) and illustrates the
estimator in several demanding settings.

The rest of the paper proceeds as follows. I begin by restating the
absorbing-treatment setup and the logic of weighted stacked DID. I then
introduce CBWSDID and show how matching and weighting refinements can be
embedded in the stacked design through first-stage design weights. Next,
I extend the estimator to repeated switch-on and switch-off episodes.
The empirical performance of the estimator is then illustrated in a
simulation design, in a stress-test application based on Trounstine
(\citeproc{ref-trounstine2020}{2020}), and in a repeated-treatment
application on democracy and growth based on Acemoglu et al.
(\citeproc{ref-acemoglu2019}{2019}).

\section{Setup}\label{setup}

\subsection{Staggered adoption and stacked
DID}\label{staggered-adoption-and-stacked-did}

Let \(s = 1, \dots, S\) index groups and \(t = 1, \dots, T\) index
calendar time. Treatment is staggered, so each group has a first
treatment time \(A_s \in \{1, \dots, T, \infty\}\) where
\(A_s = \infty\) denotes never treated. Let \(Y_{s,t}(0)\) denote the
untreated potential outcome for group \(s\) at time \(t\), and let
\(Y_{s,t}(a)\) denote the potential outcome at time \(t\) if group \(s\)
first adopts treatment in period \(a\) and remains treated thereafter.

For a cohort first treated in period \(a\) and an event time \(e=t-a\),
define the group-time average treatment effect on the treated as

\[
ATT(a, a+e) = \mathbb{E}\left[Y_{s,a+e}(a) - Y_{s,a+e}(0) \mid A_s = a\right]
\]

To estimate this effect, the stacked DID estimator proposes the
following logic. Fix a uniform event window
\(\kappa = (\kappa_{pre}, \kappa_{post})\) where \(\kappa_{pre}\) and
\(\kappa_{post}\) are desired length of the pre- and post-treatment
period. For each sub-experiment \(a\), define the treated and the
clean-control sets as

\[
\mathcal{D}_a = \{s : A_s = a\}
\qquad
\mathcal{C}_a = \{s : A_s > a + \kappa_{post}\}
\]

The clean-control definition in that case is stronger than ``not yet
treated at event time \(e\)'' since it keeps the control group fixed
across the entire event window. Let \(\Omega_\kappa\) denote the set of
(trimmed) treated cohorts \(a\) for which the full event window is
observed and the clean-control rule is feasible (so,
\(\mathcal{C}_a \neq \emptyset\)). Because of that control units may
appear more than once in the analysis. The numbers of eligible control
and treated units are

\[
N_a^D = ||\mathcal{D}_a||
\qquad
N_a^C = ||\mathcal{C}_a||
\qquad
N_{\Omega_\kappa}^D = \sum_{a \in \Omega_\kappa} N_a^D
\qquad
N_{\Omega_\kappa}^C = \sum_{a \in \Omega_\kappa} N_a^C
\] where \(N_a^D\) and \(N_a^C\) are the numbers of eligible treated and
control units for sub-experiment \(a\) respectively;
\(N_{\Omega_\kappa}^D\) and \(N_{\Omega_\kappa}^C\) are the total
numbers of eligible treated and controls units for all sub-experiment
respectively.

The target parameter is then the trimmed aggregate ATT as defined by
Wing et al. (\citeproc{ref-wing2024}{2024})

\[
\theta_\kappa^e
=
\sum_{a \in \Omega_\kappa}
ATT(a, a+e) \times \frac{N_a^D}{N_{\Omega_\kappa}^D}
\] It averages cohort-specific effects using treated-cohort shares that
are stable across event time.

For each admissible \(a \in \Omega_\kappa\), create a sub-experiment
consisting of units in \(\mathcal{D}_a \cup \mathcal{C}_a\) observed
over event times \(e \in \{-\kappa_{pre}, \dots, \kappa_{post}\}\),
where calendar time is \(t = a + e\). Then, for any event time \(e\),
define the change from the reference period \(-1\) as

\[
\Delta Y_{s,a+e} = Y_{s,a+e} - Y_{s,a-1}
\] Within sub-experiment \(a\), define the treated and control mean
changes as

\[
\bar{\Delta}_{a,e}^{D}
=
\frac{1}{N_a^D}\sum_{s \in \mathcal{D}_a}\Delta Y_{s,a+e}
\qquad
\bar{\Delta}_{a,e}^{C}
=
\frac{1}{N_a^C}\sum_{s \in \mathcal{C}_a}\Delta Y_{s,a+e}
\]

Then the cohort-specific DID is

\[
DID_{a,e} = \bar{\Delta}_{a,e}^{D} - \bar{\Delta}_{a,e}^{C}
\]

Under no anticipation and within-subexperiment parallel trends,
\(DID_{a,e}\) identifies \(ATT(a, a+e)\).

\subsection{Weighted stacked DID}\label{weighted-stacked-did}

The ordinary stacked DID estimator pools all treated observations across
sub-experiments and all clean-control observations across
sub-experiments. At event time \(e\), this pooled contrast can be
written as

\[
DID_e^{SDID}
=
\sum_{a \in \Omega_\kappa}
\frac{N_a^D}{N_{\Omega_\kappa}^D}\bar{\Delta}_{a,e}^{D}
-
\sum_{a \in \Omega_\kappa}
\frac{N_a^C}{N_{\Omega_\kappa}^C}\bar{\Delta}_{a,e}^{C}
\]

This expression reveals the central problem emphasized by Wing et al.
(\citeproc{ref-wing2024}{2024}). The treated side is aggregated using
the cohort shares \(N_a^D / N_{\Omega_\kappa}^{Treated}\), while the
control side is aggregated using a different set of shares,
\(N_a^C / N_{\Omega_\kappa}^{Control}\). Therefore, untreated trends do
not generally cancel in the pooled contrast even if parallel trends
holds within each sub-experiment.

Wing et al. (\citeproc{ref-wing2024}{2024}) correct the aggregation
problem by assigning each control observation in sub-experiment \(a\)
the sample weight

\[
Q_{sa}
=
\begin{cases}
1, & s \in \mathcal{D}_a\\[6pt]
\dfrac{N_a^D / N_{\Omega_\kappa}^D}
      {N_a^C / N_{\Omega_\kappa}^C}, & s \in \mathcal{C}_a
\end{cases}
\]

With these corrective weights, the pooled control trend is reaggregated
using the same cohort shares as the treated trend:

\[
DID_e^{WSDID}
=
\sum_{a \in \Omega_\kappa}
\frac{N_a^D}{N_{\Omega_\kappa}^D}\bar{\Delta}_{a,e}^D
-
\sum_{a \in \Omega_\kappa}
\frac{N_a^D}{N_{\Omega_\kappa}^D}\bar{\Delta}_{a,e}^C
=
\sum_{a \in \Omega_\kappa}
\frac{N_a^D}{N_{\Omega_\kappa}^D}DID_{a,e}
\]

Hence, under the usual stacked-DID assumptions, weighted stacked DID
identifies \(\theta_\kappa^e\) rather than a distorted mixture of
treated and control trends.

In practice, the estimator can be implemented by weighted least squares
on the stacked sample. Since the data is stacked, it is natural to
change the notation for unit from \(s\) (that represents unit) to \(sa\)
(that represents unit in a sub-experiment). The model specification is

\[
Y_{sa,e} 
=
\underbrace{\alpha_{sa} + \gamma_{ae}}_{\text{Fixed effects}}
+ 
\underbrace{\sum_{h = \kappa_{pre}, \ h \neq -1}^{\kappa_{post}}\beta_h D_{sa} \mathbf{1}\{e = h\}}_{\text{Dynamic effects}}
+ \varepsilon_{sa, e}
\]

where \(\alpha_{sa}\) and \(\gamma_{ae}\) are sub-experiment fixed
effects and \(D_{sa}\) is a time-invariant indicator of treatment in
sub-experiment \(a\). Observations are weighted by \(Q_{sa}\) and
coefficients \(\beta_e\) are the trimmed aggregate ATT at event time
\(e\). The variance estimation is based on the assumption that
observations are dependent within units \(s\) but independent across
them. Thus, cluster-robust variance estimators are usually used, while
clustering on \(sa\) is also possible. However, as Wing et al.
(\citeproc{ref-wing2024}{2024}) show, they lead to the same desired
rejection rates (and both suffer from small number of clusters), while
the first option should be preferred to additionally account for
repeated observations across sub-experiments.

\section{Covariate-Balanced Weighted Stacked
DID}\label{covariate-balanced-weighted-stacked-did}

\subsection{Design weights and final stacked
weights}\label{design-weights-and-final-stacked-weights}

The previous setup assumes that the clean-control mean
\(\bar{\Delta}_{a,e}^C\) is already a credible estimate of the untreated
trend for the treated cohort in sub-experiment \(a\). Often it is
implausible, because the clean-control set may differ from treated units
in observed covariates, lagged outcomes, or both. To address this
problem, one can add a within-subexperiment design stage before the
corrective aggregation stage.

For each admissible cohort \(a \in \Omega_\kappa\), let \(X_{sa}\)
denote a vector of pre-treatment characteristics constructed only from
information at \(t \le a-1\). It may include baseline covariates, lagged
outcomes, pre-trend summaries etc. Using the sample
\(\mathcal{D}_a \cup \mathcal{C}_a\), construct nonnegative design
weights \(b_{sa}\) for control units \(s \in \mathcal{C}_a\). Thus,
treated units are left unchanged, so they continue to receive unit
weight, and the weighted control mean change within sub-experiment \(a\)
is

\[
\bar{\Delta}_{a,e}^{C,b}
=
\frac{\sum_{s \in \mathcal{C}_a} b_{sa}\Delta Y_{s,a+e}}
     {\sum_{s \in \mathcal{C}_a} b_{sa}}
\]

Therefore, the covariate-balanced cohort-specific DID is

\[
DID_{a,e}^{\,b}
=
\bar{\Delta}_{a,e}^D - \bar{\Delta}_{a,e}^{C,b}
\]

It has a clear design-based interpretation. The weights \(b_{sa}\)
choose which clean controls matter for approximating the untreated path
of cohort \(a\) and solve a within-subexperiment comparability problem.
However, they do not solve the across-subexperiment aggregation problem
and we need to incorporate Wing et al.'s corrective weights. It requires
a new definition of control group's number of observation. Define the
effective control mass in sub-experiment \(a\) as

\[
\widetilde N_a^C = \sum_{s \in \mathcal{C}_a} b_{sa}
\qquad
\widetilde N_{\Omega_\kappa}^C = \sum_{a \in \Omega_\kappa}\widetilde N_a^C
\]

Because treated cohorts are not reweighted (while it is theoretically
possible), \(\widetilde N_a^D = N_a^D\) and
\(\widetilde N_{\Omega_\kappa}^D = N_{\Omega_\kappa}^D\). The final
sample weights are therefore

\[
W_{sa}
=
\begin{cases}
1, & s \in \mathcal{D}_a,\\[6pt]
b_{sa}\times
\dfrac{N_a^D / N_{\Omega_\kappa}^D}
      {\widetilde N_a^C / \widetilde N_{\Omega_\kappa}^C}, & s \in \mathcal{C}_a
\end{cases}
\]

Note, it is not necessary to normalize \(b_{sa}\) to sum to one within
each sub-experiment, because the weighted control mean
\(\bar{\Delta}_{a,e}^{C,b}\) is already normalized by
\(\sum_{s \in \mathcal{C}_a} b_{sa}\). So, replacing \(b_{sa}\) by
\(c_a b_{sa}\) for any positive constant \(c_a\) leaves the
within-subexperiment DID unchanged. Additionally, across sub-experiments
the role of normalization is played by the effective control mass
\(\widetilde N_a^C\) and the corrective stacked factor below. In
particular,

\[
\sum_{s \in \mathcal{C}_a} W_{sa}
=
\widetilde N_{\Omega_\kappa}^C \times \frac{N_a^D}{N_{\Omega_\kappa}^D}
\]

so the total weighted control mass in sub-experiment \(a\) is
automatically proportional to the treated-cohort share, regardless of
the raw scale of \(b_{sa}\).

Equivalently, the estimator can be written directly as the treated-share
average of within-subexperiment balanced DID contrasts:

\[
DID_e^{CBWSDID}
=
\sum_{a \in \Omega_\kappa}
\frac{N_a^D}{N_{\Omega_\kappa}^D}
DID_{a,e}^{\,b}
\]

An attractive feature of the specification above is that it leaves the
treated side of each cohort unchanged. As long as the full set of
treated cohorts in \(\Omega_\kappa\) is retained, the target parameter
remains

\[
\theta_\kappa^e
=
\sum_{a \in \Omega_\kappa}
ATT(a, a+e)\times \frac{N_a^D}{N_{\Omega_\kappa}^D}
\]

In other words, proposed estimator only changes the way untreated
counterfactual trends are estimated, but it does not change the cohort
weights in the aggregate ATT. More aggressive preprocessing steps that
discard treated units or entire treated cohorts may still be useful, but
they change the estimand to a something like overlap-trimmed ATT.

How should the design weights \(b_{sa}\) be constructed? By matching of
weighting that have become widely used (see
\citeproc{ref-stuart2010}{Stuart 2010}; \citeproc{ref-austin2015}{Austin
and Stuart 2015}). Fortunately, both approaches are theoratically
equivalent for proposed estimator, because they produce a nonnegative
control weight \(b_{sa}\) for each unit \(s \in \mathcal{C}_a\) in
sub-experiment \(a\). In the absence of covariate adjustment,
\(b_{sa}=1\) for all clean controls, which returns the weighted stacked
DID estimator of Wing et al. (\citeproc{ref-wing2024}{2024}). Under
matching-based refinement, the resulting weights depend on the specific
method. For discrete matching procedures (such as nearest-neighbor
matching without replacement) \(b_{sa}\) is typically binary: matched
controls receive weight one and unmatched controls receive weight zero.
Under matching with replacement or other designs in which the same
control can be reused multiple times, \(b_{sa}\) can instead take
integer values that record how often a control appears in the matched
sample, or normalized versions of such counts. Other matching designs,
such as optimal or full matching, can likewise be represented through
nonnegative design weights. Under weighting-based refinement, \(b_{sa}\)
is a continuous balancing weight obtained from inverse-probability
weighting, entropy balancing, overlap weighting, CBPS, or related
methods. Thus, the proposed estimator does not depend on a particular
refinement method.

\subsection{Identification
assumptions}\label{identification-assumptions}

The proposed estimator replaces unconditional parallel trends with a
weighted version that is specific to each sub-experiment. Thus, the
identifying assumptions are similar to those of the stacked DID
estimator. Specifically,

\textbf{Assumption 1 (No anticipation).} For every admissible cohort
\(a \in \Omega_\kappa\),

\[
ATT(a,a+e) = 0 \qquad \text{for all } e < 0
\]

\textbf{Assumption 2 (Within-subexperiment weighted parallel trends).}
For every \(a \in \Omega_\kappa\) and every
\(e \in \{-\kappa_{pre},\dots,\kappa_{post}\}\),

\[
\mathbb{E}\left[Y_{s,a+e}(0)-Y_{s,a-1}(0)\mid A_s=a\right]
=
\mathbb{E}_{b_a}\left[Y_{s,a+e}(0)-Y_{s,a-1}(0)\mid s \in \mathcal{C}_a\right]
\]

\textbf{Assumption 3 (Overlap and nondegeneracy).} For every
\(a \in \Omega_\kappa\), the treated and clean-control sets are nonempty
and the effective control mass satisfies
\(0 < \widetilde N_a^C < \infty\).

\textbf{Assumption 4 (Pre-treatment refinement).} The design weights
\(b_{sa}\) are constructed only from information dated before \(a-1\).
So, no post-treatment bias is introduced.

Under Assumptions 1-4,

\[
DID_{a,e}^{\,b} = ATT(a,a+e)
\qquad
\text{for each } a \in \Omega_\kappa \text{ and admissible } e
\]

and therefore

\[
DID_e^{CBWSDID} = \theta_\kappa^e
\]

Equivalently, the coefficient \(\beta_e\) estimated by the weighted
stacked event-study regression with weights \(W_{sa}\) identifies the
trimmed aggregate ATT at event time \(e\).

The logic is straightforward. Assumptions 1-4 guarantee that the
within-subexperiment balanced DID identifies the cohort-specific ATT.
The corrective factor in \(W_{sa}\) then aligns the weighted control
contribution with the treated-cohort shares
\(N_a^D/N_{\Omega_\kappa}^D\), so the pooled estimator aggregates the
cohort-specific effects using exactly the target weights. In this sense,
the proposed estimator combines the covariate-adjustment logic of
Callaway and Sant'Anna (\citeproc{ref-callaway2021}{2021}) within
sub-experiments and the aggregation logic of Wing et al.
(\citeproc{ref-wing2024}{2024}) across sub-experiments.

\section{\texorpdfstring{Extension to Repeated \(0 \to 1\)
Cases}{Extension to Repeated 0 \textbackslash to 1 Cases}}\label{extension-to-repeated-0-to-1-cases}

\subsection{Reformulation of absorbing
design}\label{reformulation-of-absorbing-design}

Let \(D_{s,t} \in \{0,1\}\) denote a binary treatment that may turn on,
turn off, and later turn on again. Fix the same event-study window
\(\kappa = (\kappa_{pre}, \kappa_{post})\) as before, and introduce a
history window of length \(L \geq 1\). For each unit \(s\) and calendar
time \(\tau\), define the recent treatment history

\[
H_{s,\tau}^{(L)} = (D_{s,\tau-L}, \dots, D_{s,\tau-1}) \in \{0,1\}^L
\] Next, for a given history profile \(h\) and switch time \(\tau\),
define the treated episode set as

\[
\mathcal{D}_{\tau,h}^{01}
=
\left\{
s :
H_{s\tau}^{(L)} = h,\;
D_{s,\tau-1}=0,\;
D_{s,\tau+r}=1 \text{ for all } r=0,\dots,\kappa_{post}
\right\}
\] Units in \(\mathcal{D}_{\tau,h}^{01}\) switch on at time \(\tau\) and
remain on throughout the post-treatment portion of the event window. The
role of \(H_{s,\tau}^{(L)}\) is to control for how much recent treatment
history must be held fixed when constructing comparable switch-on
episodes. Basically, the idea is the same as in PanelMatch
(\citeproc{ref-imai2023}{Imai et al. 2023}) where treated observations
are matched to controls with identical treatment histories over a
prespecified number of lags before treatment. Thus, \(L\) is different
from the \(\kappa_{pre}\). The later controls which pre-treatment and
post-treatment effects are displayed, whereas \(L\) controls how much
recent treatment history must be held fixed when constructing comparable
switch-on episodes. Admissible control episodes must remain untreated
throughout the window. Thus, treatment reversals may occur elsewhere in
the panel and may appear in the lag-history vector \(H_{s,\tau}^{(L)}\),
but not in \(\{-\kappa_{pre}, \dots, \kappa_{post}\}\). The
corresponding stable untreated control episodes are

\[
\mathcal{C}_{\tau,h}^{0}
=
\left\{
s :
H_{s\tau}^{(L)} = h,\;
D_{s,\tau+r}=0 \text{ for all } r=0,\dots,\kappa_{post}
\right\}
\] The control set is indexed by stable untreated episodes rather than
by never-treated units. This is deliberate: in a non-absorbing design,
what matters is not whether a unit is never treated in its entire
lifetime, but whether it remains untreated and comparable over the local
event window.

To make this extension operational, one needs a finite-memory
restriction. To simplify, for event times in the window of interest,
potential outcomes should depend on past treatment history only through
the most recent \(L\) periods and through treatment status over the
current event window. Under that restriction, a convenient treatment
effect is

\[
ATT^{01}(\tau,h,\tau+e)
=
\mathbb{E}\left[
Y_{s,\tau+e}(h \to 1) - Y_{s,\tau+e}(h \to 0)
\mid s \in \mathcal{D}_{\tau,h}^{01}
\right]
\]

where \(Y_{s,\tau+e}(h \to 1)\) denotes the potential outcome at event
time \(e\) for a unit with recent history \(h\) that switches on at
\(\tau\) and remains treated over the event window, and
\(Y_{s,\tau+e}(h \to 0)\) denotes the analogous potential outcome for a
unit with the same recent history that remains untreated over the event
window. This formulation preserves the spirit of the absorbing-treatment
setup while allowing the same unit to contribute multiple admissible
episodes at different calendar times. It also makes clear why recent
treatment history must be part of the design stage rather than an
afterthought.

The natural extension is to analyze not only treatment adoption
(\(0 \to 1\)), but also treatment reversals (\(1 \to 0\)). This can be
done by mechanically making \(D^{1\to0}=1-D^{0\to1}\). So, in terms of
estimation nothing changes.

\subsection{Episode-weighted estimand and stacked
weights}\label{episode-weighted-estimand-and-stacked-weights}

Let \(\Omega_{L,\kappa}^{01}\) denote the set of admissible \((\tau,h)\)
episode types for which the full event window is observed and both
treated and stable untreated episodes exist. Define

\[
M_{\tau,h}^D = ||\mathcal{D}_{\tau,h}^{01}||
\qquad
M_{\tau,h}^C = ||\mathcal{C}_{\tau,h}^{0}||
\qquad
M_{\Omega}^{D}
=
\sum_{(\tau,h)\in\Omega_{L,\kappa}^{01}} M_{\tau,h}^D
\]

The natural analogue of the trimmed aggregate ATT is then

\[
\theta_{e}^{01}(L,\kappa)
=
\sum_{(\tau,h)\in\Omega_{L,\kappa}^{01}}
ATT^{01}(\tau,h,\tau+e)\times \frac{M_{\tau,h}^D}{M_{\Omega}^{D}}
\]

This is episode-weighted rather than unit-weighted parameter. A unit
that contributes multiple admissible \(0 \to 1\) episodes receives more
total weight than a unit that contributes only one episode. Accordingly,
\(\theta_{e}^{01}(L,\kappa)\) answers the question ``what is the average
effect of an admissible switch-on episode?'' rather than ``what is the
average effect for a switching unit?'' A unit-weighted alternative can
be defined, but it would require an additional unit-level re-aggregation
step and is not developed here.

Within each admissible episode type \((\tau,h)\), one can construct
control-only design weights \(b_{s,\tau,h}\) as in case of staggered
treatment adoption described above. Let

\[
\widetilde M_{\tau,h}^{C}
=
\sum_{s \in \mathcal{C}_{\tau,h}^{0}} b_{s,\tau,h}
\qquad
\widetilde M_{\Omega}^{C}
=
\sum_{(\tau,h)\in\Omega_{L,\kappa}^{01}}\widetilde M_{\tau,h}^{C}
\]

Then the direct analogue of the earlier stacked weight is

\[
W_{s,\tau,h}^{01}
=
\begin{cases}
1, & s \in \mathcal{D}_{\tau,h}^{01},\\[6pt]
b_{s,\tau,h}\times
\dfrac{M_{\tau,h}^D / M_{\Omega}^D}
      {\widetilde M_{\tau,h}^C / \widetilde M_{\Omega}^C},
& s \in \mathcal{C}_{\tau,h}^{0}
\end{cases}
\]

Define the weighted control mean change within episode type \((\tau,h)\)
by

\[
\bar{\Delta}_{\tau,h,e}^{C,b}
=
\frac{\sum_{s \in \mathcal{C}_{\tau,h}^{0}} b_{s,\tau,h}\left(Y_{s,\tau+e}-Y_{s,\tau-1}\right)}
     {\sum_{s \in \mathcal{C}_{\tau,h}^{0}} b_{s,\tau,h}}
\]

and define the treated mean change \(\bar{\Delta}_{\tau,h,e}^{D}\)
analogously over \(\mathcal{D}_{\tau,h}^{01}\). The pooled estimator can
then be written as

\[
DID_{e}^{01,CBWSDID}
=
\sum_{(\tau,h)\in\Omega_{L,\kappa}^{01}}
\frac{M_{\tau,h}^D}{M_{\Omega}^{D}}
\left(
\bar{\Delta}_{\tau,h,e}^{D} - \bar{\Delta}_{\tau,h,e}^{C,b}
\right)
\]

Conceptually, nothing changes relative to the absorbing-treatment case.
The design weights \(b_{s,\tau,h}\) solve the within-episode
comparability problem, while the corrective factor aligns control
episodes with the treated-episode shares that define the target
parameter. Thus, the same weighted regression model can be used. Note,
because the same unit may contribute multiple admissible episodes over
time, dependence across episodes becomes even more salient in this
extension, which further supports unit-level clustering.

\subsection{Identification
assumptions}\label{identification-assumptions-1}

The identifying assumptions for the repeated-treatment extension
parallel Assumptions 1-4, but the conditioning object changes from a
first-adoption cohort \(a\) to an episode type \((\tau,h)\), and a
finite-memory restriction must be added. Specifically,

\textbf{Assumption R1 (Finite memory).} For event times in the window of
interest \(\{-\kappa_{pre},\dots,\kappa_{post}\}\), the relevant
potential outcomes depend on the pre-\(\tau\) treatment path only
through the recent history \(H_{s\tau}^{(L)}\).

\textbf{Assumption R2 (No anticipation of switch-on episodes).} For
every admissible \((\tau,h)\in\Omega_{L,\kappa}^{01}\),

\[
\mathbb{E}\left[
Y_{s,\tau+e}(h \to 1) - Y_{s,\tau+e}(h \to 0)
\mid s \in \mathcal{D}_{\tau,h}^{01}
\right]
= 0
\qquad
\text{for all } e < 0
\]

\textbf{Assumption R3 (Stable episode design).} Treated episodes in
\(\mathcal{D}_{\tau,h}^{01}\) remain treated throughout the
post-treatment event window, and control episodes in
\(\mathcal{C}_{\tau,h}^{0}\) remain untreated throughout that same
window. Note, this assumption can be relaxed if treated episodes are
defined only by treatment onset at \(e=0\) and may subsequently reverse
within the event window. This is similar to allowing treatment reversal
in the PanelMatch estimator (\citeproc{ref-imai2023}{Imai et al. 2023}).

\textbf{Assumption R4 (History-conditional weighted parallel trends).}
For every admissible \((\tau,h)\in\Omega_{L,\kappa}^{01}\) and
admissible event time \(e\),

\[
\mathbb{E}\left[
Y_{s,\tau+e}(h \to 0) - Y_{s,\tau-1}(h \to 0)
\mid s \in \mathcal{D}_{\tau,h}^{01}
\right]
=
\mathbb{E}_{b_{\tau,h}}\left[
Y_{s,\tau+e}(h \to 0) - Y_{s,\tau-1}(h \to 0)
\mid s \in \mathcal{C}_{\tau,h}^{0}
\right]
\]

\textbf{Assumption R5 (Episode-level overlap and nondegeneracy).} For
every admissible \((\tau,h)\in\Omega_{L,\kappa}^{01}\), the treated and
stable untreated episode sets are nonempty and
\(0 < \widetilde M_{\tau,h}^C < \infty\).

\textbf{Assumption R6 (Pre-treatment design and episode invariance).}
The design weights \(b_{s,\tau,h}\) are constructed only from
information dated before \(\tau-1\), and for a fixed admissible episode
type \((\tau,h)\) they are held fixed across all event times in the
window.

Similarly to absorbing staggered-adoption design, under Assumptions
R1-R6,

\[
\bar{\Delta}_{\tau,h,e}^{D} - \bar{\Delta}_{\tau,h,e}^{C,b}
=
ATT^{01}(\tau,h,\tau+e)
\qquad
\text{for each admissible } (\tau,h) \text{ and } e
\]

and therefore

\[
DID_{e}^{01,CBWSDID} = \theta_{e}^{01}(L,\kappa)
\]

Thus, the repeated-treatment extension preserves the same two-step
structure as the absorbing-treatment design: design adjustment
identifies episode-specific counterfactuals, and corrective stacked
weighting recovers the desired episode-weighted aggregate.

Note, in the baseline absorbing staggered-adoption design, the usual
distinction is between never-treated and not-yet-treated controls
(\citeproc{ref-callaway2021}{Callaway and Sant'Anna 2021}). In the
repeated-treatment design, however, that is no longer the most helpful
distinction. The relevant comparison is between switch-on episodes and
stable untreated episodes with the same recent treatment history. This
has an important implication: a never-treated unit has an all-zero
treatment history in every lag window. Therefore, under exact or
near-exact matching on recent treatment history, never-treated units are
admissible controls only for switch-on episodes whose recent history is
also all zeros. So, they are not generally valid controls for episodes
that occur after earlier exposures to treatment. By contrast, later
switchers and previously treated units can serve as controls if, at time
\(\tau\), they are untreated throughout the event window and share the
relevant recent history profile.

\section{Inference and Variance
Estimation}\label{inference-and-variance-estimation}

The proposed estimator has two stages. The design stage constructs
either matched samples or balancing weights within each sub-experiment
or episode type, and the second stage estimates a weighted stacked DID.
In both the absorbing-treatment and repeated-treatment cases, the same
unit may appear multiple times in the stacked sample, so dependence
across observations is naturally handled by clustering at the unit
level.

The simplest approach is to condition on the estimated design weights
and treat the final weights \(W_{sa}\) or \(W_{s,\tau,h}^{01}\) as fixed
in the second-stage regression. This yields a cluster-robust variance
estimator for the weighted stacked regression conditional on the
estimated design. This approach is close to the PanelMatch estimator,
where standard errors are often computed conditional on the weights
implied by the matching step rather than by treating the entire matching
procedure as part of the stochastic expansion
(\citeproc{ref-imai2023}{Imai et al. 2023}).

Another option is to use some sort of bootstrap. The simplest option is
to use a cluster bootstrap conditional on estimated design weights.
Resample units or groups, keep the estimated first-stage matching or
weighting structure fixed, and recompute only the second-stage stacked
estimator. Other option is to use a cluster bootstrap with re-estimation
of smooth balancing weights. When the first stage is smooth (as with
balancing-weight estimators), a more ambitious bootstrap can rebuild the
first-stage design weights in each replication. This more fully
propagates first-stage estimation uncertainty. However, the main caveat
concerns nonsmooth matching estimators, because a standard bootstrap
that recomputes nearest-neighbor or similar techniques in each
replication is not generally valid (see \citeproc{ref-abadie2008}{Abadie
and Imbens 2008}).

\section{Simulation example}\label{simulation-example}

To illustrate the finite-sample behavior of the estimator in a
controlled setting, consider a balanced panel with \(S=500\) units
observed annually from 2000 through 2012. Treatment is absorbing. Each
unit belongs to one of five treatment-timing groups: first adoption in
2004, 2005, 2006, or 2007, or never treated. Adoption timing is
correlated with observed covariates, so treatment selection is not
random across cohorts.

Each unit \(s\) is assigned a continuous covariate \(x_{1s}\), a binary
covariate \(x_{2s}\), a unit intercept \(\alpha_s\), and a unit-specific
untreated slope \(\delta_s\). The slope is constructed so that the same
covariates that predict treatment timing also shift untreated trends:

\[
\delta_s = 0.07x_{1s} + 0.05x_{2s} + \eta_s,
\qquad
\eta_s \sim \mathcal{N}(0,0.03^2)
\]

Treatment timing is then drawn from a multinomial assignment rule over
the support \(\{2004,2005,2006,2007,\infty\}\), with relative cohort
scores that depend on \((x_{1s},x_{2s})\). As a result, units with
different treatment timings also differ systematically in observed
characteristics and in their untreated outcome trajectories. Evolution
of treatment adoption is plotted in Figure~\ref{fig-d} below.

\begin{figure}

\centering{

\pandocbounded{\includegraphics[keepaspectratio]{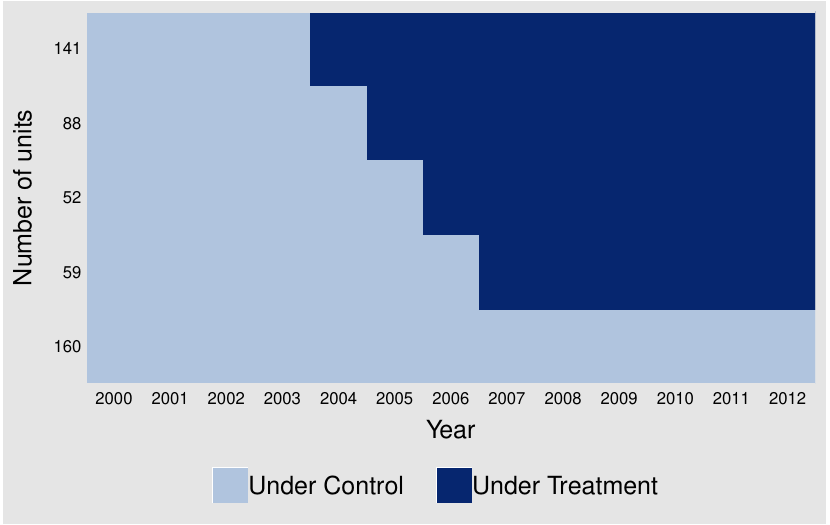}}

}

\caption{\label{fig-d}Treatment adoption map in absorbing-treatment
simulation.}

\end{figure}%

Untreated outcomes evolve accordingly to a dynamic panel process with
unit heterogeneity, common time effects, and covariate-dependent slopes:

\[
Y_{st}(0)
=
0.45\,Y_{s,t-1}(0)
+ \alpha_s
+ \lambda_t
+ 0.50x_{1s}
+ 0.35x_{2s}
+ \delta_s(t-1)
+ \varepsilon_{st},
\]

where \(\lambda_t\) is a smooth calendar-time effect and
\(\varepsilon_{st}\) is a random shock. This data-generating process
deliberately violates unconditional parallel trends because the same
observed covariates affect both treatment timing and untreated trend
growth.

Treatment effects are dynamic and event-time specific. The treatment
effect path is set to zero for all pre-treatment event times, equal to
\(-0.40\) at treatment onset, equal to \(-0.80\) one period after
adoption, and equal to \(-1.10\) for all subsequent post-treatment
periods in the simulation window. The observed outcome is therefore

\[
Y_{st} = Y_{st}(0) + \tau_e,
\]

This simulation is designed to highlight exactly the environment that
motivates the proposed estimator: untreated trends differ systematically
across treatment cohorts, making ordinary stacked DID (and its weighted
version too) vulnerable to bias, while covariate balancing within
sub-experiments and corrective weighting across sub-experiments are both
relevant.

Estimators under considerations are ordinary and weighted stacked DID as
well as CBWSDID with matching or weighting refinement. The former
version is based on nearest neighbor matching with replacement and
Mahalanobis distance when treated-to-controls ratio is four. The latter
version is done with entropy balancing
(\citeproc{ref-hainmueller2012}{Hainmueller 2012}). In both versions the
pre-treatment covariates for refinement \(X_{sa}\) include three lags of
\(Y_{st}\) and \(x_{1st}\) as well as exact matching on \(x_{2s}\),
which is time-invariant binary variable. The event-study window is
\(\kappa=(-3,2)\).

\begin{figure}

\centering{

\pandocbounded{\includegraphics[keepaspectratio]{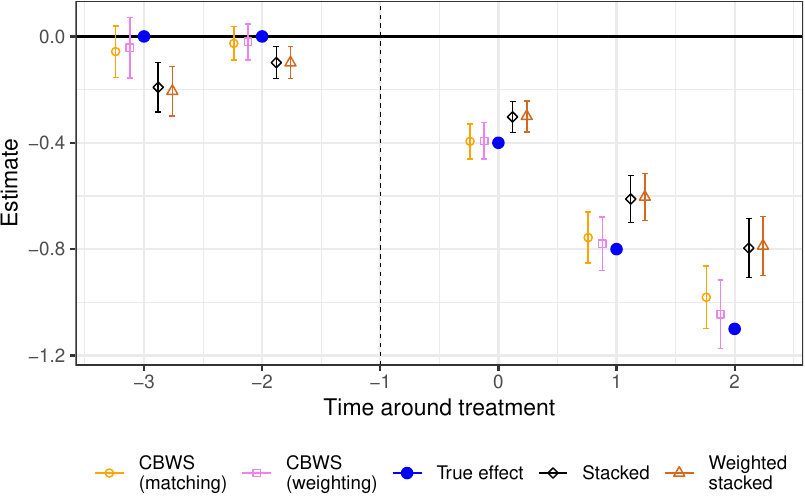}}

}

\caption{\label{fig-sim}Event-study estimates in the absorbing-treatment
simulation.}

\end{figure}

The simulation results are presented in Figure~\ref{fig-sim}. It is seen
that ordinary stacked DID and weighted stacked DID exhibit noticeable
negative pre-trends and significant bias in post-treatment effects
relative to the true dynamic path (big beautiful blue point). In
contrast, both covariate-balanced estimators substantially reduce the
spurious pre-treatment coefficients and move the post-treatment
estimates closer to the true treatment-effect profile. In this
simulation, the weighting-based CBWSDID specification comes closest to
the true post-treatment path, while the matching-based version also
improves markedly over the two stacked benchmarks.

To compare estimators systematically, I did a Monte Carlo study
repeating aformentioned simulation 5000 times. The results are presented
in Table~\ref{tbl-sim-mc}. It is seen that stacked DID and weighted
stacked DID continue to exhibit substantial bias both in the last
pre-treatment period and in the post-treatment periods, and accordingly
they reject the true effect at very high rates. By contrast, both
versions of CBWSDID are much closer to the true dynamic path. The
weighting-based version performs especially well in this DGP: its
average estimated effects are very close to the truth and the rejection
rate of the true effect remains low across the reported post-treatment
horizons. The matching-based version still improves markedly on the two
stacked benchmarks, but its finite-sample performance is somewhat worse
than that of the weighting-based version. Nevertheless, this comparison
of matching and weighting should not be overstated. Matching estimators
are often more sensitive to practical tuning choices, such as the number
of controls per treated unit, replacement, calipers, or the distance
metric. In applied work, researchers often iterate over such choices to
obtain better design balance, so the matching results reported here
should be interpreted as results for one particular implementation (and
the author's laziness in selecting a better specification) rather than a
general conclusion. Weighting is in that sense often less
tuning-intensive, which makes it look attractive in simulation. At the
same time, in many empirical applications weighting may behave worse
than matching because treated and control units are too dissimilar and
the resulting balancing weights become unstable or extreme. Thus, the
main lesson from Table~\ref{tbl-sim-mc} is not that weighting always
dominates matching, but that both forms of first-stage refinement
substantially improve on unrefined stacked DID in a setting where
unconditional parallel trends are implausible.

\begin{table}

\caption{\label{tbl-sim-mc}Monte Carlo performance in the
absorbing-treatment simulation.}

\centering{

\centering
\begin{tabular}{>{}l>{}c>{}c>{}c>{}c}
\toprule
\multicolumn{1}{c}{Estimator} & \multicolumn{1}{c}{\makecell[c]{Event\\time}} & \multicolumn{1}{c}{\makecell[c]{True\\effect}} & \multicolumn{1}{c}{\makecell[c]{Average estimated\\effect}} & \multicolumn{1}{c}{\makecell[c]{Rejection rate\\of true effect}}\\
\midrule
CBWSDID (matching) & -2 & 0.0 & -0.026 & 0.007\\
CBWSDID (weighting) & -2 & 0.0 & -0.008 & 0.000\\
Stacked DID & -2 & 0.0 & -0.089 & 0.882\\
Weighted stacked DID & -2 & 0.0 & -0.089 & 0.889\\
\midrule\\
CBWSDID (matching) & 0 & -0.4 & -0.377 & 0.101\\
CBWSDID (weighting) & 0 & -0.4 & -0.393 & 0.050\\
Stacked DID & 0 & -0.4 & -0.320 & 0.818\\
Weighted stacked DID & 0 & -0.4 & -0.320 & 0.823\\
\midrule\\
CBWSDID (matching) & 1 & -0.8 & -0.755 & 0.143\\
CBWSDID (weighting) & 1 & -0.8 & -0.787 & 0.040\\
Stacked DID & 1 & -0.8 & -0.645 & 0.968\\
Weighted stacked DID & 1 & -0.8 & -0.643 & 0.967\\
\midrule\\
CBWSDID (matching) & 2 & -1.1 & -1.033 & 0.183\\
CBWSDID (weighting) & 2 & -1.1 & -1.080 & 0.033\\
Stacked DID & 2 & -1.1 & -0.871 & 0.990\\
Weighted stacked DID & 2 & -1.1 & -0.869 & 0.991\\
\bottomrule
\end{tabular}

}

\end{table}%

\section{Empirical examples}\label{empirical-examples}

\subsection{The whiteness of the Fair Housing
Act}\label{the-whiteness-of-the-fair-housing-act}

\begin{figure}

\centering{

\pandocbounded{\includegraphics[keepaspectratio]{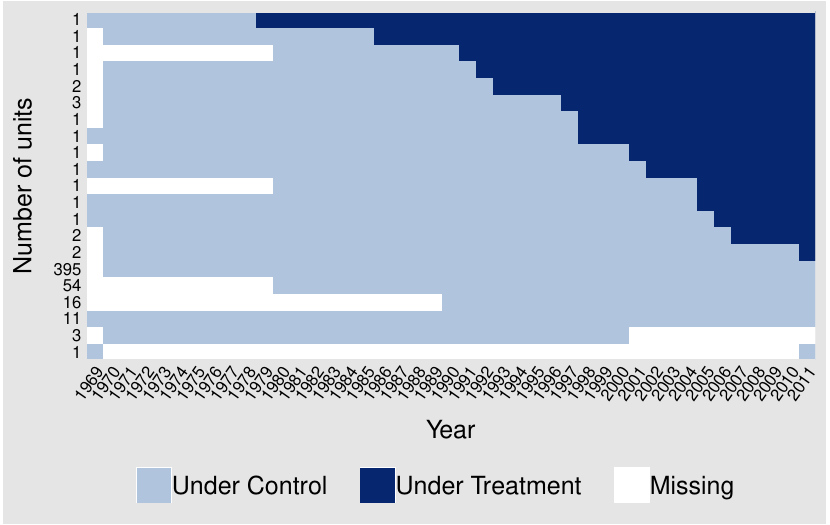}}

}

\caption{\label{fig-d-ineq}Treatment adoption map in Trounstine (2020).}

\end{figure}

As a first empirical illustration, I use the study by Trounstine
(\citeproc{ref-trounstine2020}{2020}) who studied how Fair Housing Act
adoption affects racial segregation measured as city whiteness at the
city-year level. Treatment is a standard staggered-adoption policy, so
it fits directly into the absorbing-treatment version of the estimator.
Meanwhile, the setting is empirically demanding because treated and
untreated cities exhibit visibly different pre-treatment dynamics, which
makes it a good stress test for design-based refinement. The following
analysis uses the replication data from Trounstine
(\citeproc{ref-trounstine2020}{2020}) (obtained via replication
materials of large re-analysis study by Chiu et al.
(\citeproc{ref-chiu2025}{2025})). The map of treatment adoption is
presented in Figure~\ref{fig-d-ineq}.

Let the event-study window for this example be \(\kappa=(-10,5)\). Thus,
admissible treated cohorts are policy adoption events for which ten
pre-treatment periods and five post-treatment periods are observed.
Clean controls are cities that remain untreated throughout the
corresponding post-treatment window, so the composition of the control
group is held fixed within each sub-experiment.

Estimators under consideration are weighted stacked DID, CBWSDID with
matching or weighting refinement, TWFE, and the Sun and Abraham
(\citeproc{ref-sun2021}{2021}) estimator.\footnote{I also tried to
  include Callaway and Sant'Anna (\citeproc{ref-callaway2021}{2021})
  estimator using \texttt{did} package as an additional benchmark.
  However, for this application the package does not return stable
  estimates. The likely reason is the combination of a highly unbalanced
  panel and a large number of very small treated cohorts after balancing
  and support restrictions are imposed. In turn, \texttt{sunab} function
  from \texttt{fixest} package that implements Sun and Abraham
  (\citeproc{ref-sun2021}{2021}) estimator works well (while much slower
  than CBWSDID).} Matching refinement is based on nearest neighbor
matching with replacement and robust Mahalanobis distance (see
\citeproc{ref-rosenbaum2020}{Rosenbaum 2020, 210--11}) when
treated-to-controls ratio is 10. Weighting refinement is done via
entropy balancing. For both refinement strategies covariate
specification is the same: two lags of outcome and one lag logarithm of
population and average home prices. While original study involves much
more covariates, I use this minimal set for two reasons. Firstly, it
remains example simple and transparent; secondly, addition of new
variables does not change the results. The TWFE model is added as the
simplest estimator for comparison. In turn, the Sun-Abraham model is an
interaction-weighted estimator in the class of heterogeneity-robust DID
estimators that is suited for staggered treatment adoption and corrects
known TWFE problems (for a very brief introduction see
\citeproc{ref-chiu2025}{Chiu et al. 2025}). For both models I do not
include additional covariates such as population and housing prices
since they are plausibly affected by Fair Housing Act. In practice,
however, adding these covariates leaves the results essentially
unchanged.

\begin{figure}

\centering{

\pandocbounded{\includegraphics[keepaspectratio]{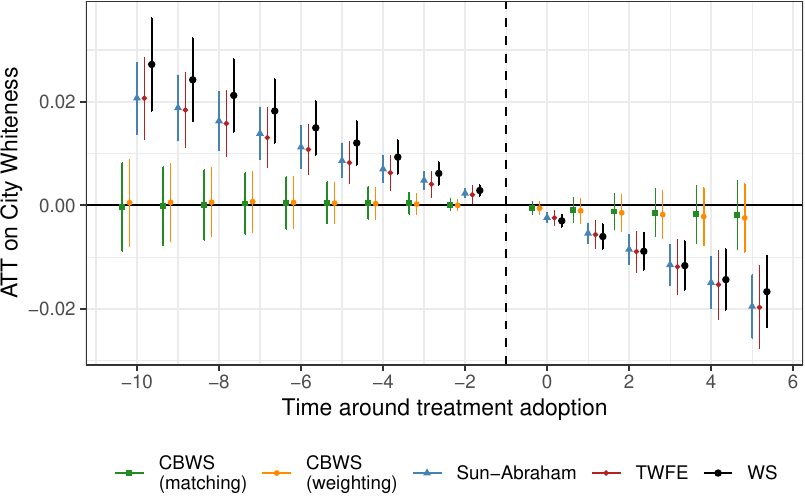}}

}

\caption{\label{fig-ineq-dynamic}Event-study comparison of estimators in
the replication of Trounstine (2020).}

\end{figure}

The results of replication using different estimators are presented in
Figure~\ref{fig-ineq-dynamic}. It is seen that all models without
refinement -- TWFE, Sun-Abraham, and weighted stacked DID -- show large
and statistically significant positive coefficients in the pre-treatment
periods, followed by a pronounced decline in city whiteness after
adoption. In other words, all three estimators recover broadly the same
substantive pattern on the raw donor pool, but they do so in a design
environment with clearly non-flat pre-trends. This is exactly the kind
of empirical setting in which a design-based refinement step is most
valuable. By contrast, both versions of CBWSDID substantially flatten
the pre-treatment path. In the matching-based and weighting-based
versions, the lead coefficients are numerically very close to zero and
their confidence intervals cover zero throughout the pre-period. At the
same time, the post-treatment decline becomes much smaller in magnitude
and is no longer statistically distinguishable from zero at any
reasonable significance level. Thus, the main role of refinement in this
application is not to sharpen an already credible design, but rather to
show that the large negative post-treatment pattern in the unrefined
estimators is tightly linked to poor treated-control comparability and
implausibility of unconditional parallel trends assumption.

\subsection{Autocracy does cause
downturn}\label{autocracy-does-cause-downturn}

\begin{figure}

\centering{

\pandocbounded{\includegraphics[keepaspectratio]{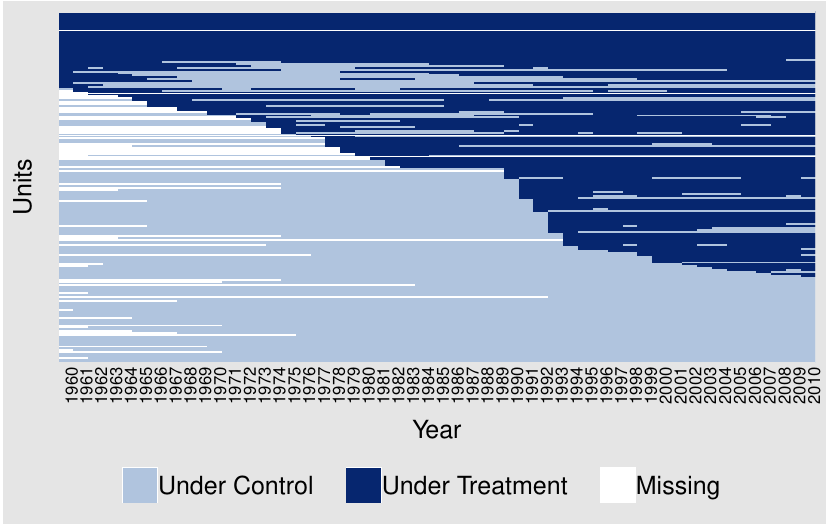}}

}

\caption{\label{fig-d-acem}Treatment adoption map in Acemoglu et
al.~(2019).}

\end{figure}

As an empirical illustration of repeated \(0\to 1\) design, I use the
setting of democracy and growth from the study by Acemoglu et al.
(\citeproc{ref-acemoglu2019}{2019}), which was also used as an example
in Imai et al. (\citeproc{ref-imai2023}{2023}) introducing the
PanelMatch estimator. The research question is simple: does democracy
cause growth? Putting differently, what is the effect of switching from
autocracy to democracy on a country's GDP per capita? The following
analysis uses the full replication panel from the study, which contains
184 countries observed annually from 1960 through 2010. The outcome is
log GDP per capita and treatment is the binary democracy indicator.
Countries may democratize, autocratize, and later democratize again, as
shown in Figure~\ref{fig-d-acem}.

The event-study window is set to \(\kappa = (-4,10)\), and recent
treatment history is defined over \(L=4\) lags. Thus, a treated episode
is a country-year observation in which democracy turns on after four
years of observed lag history and then remains on throughout the
ten-year post-treatment window. Admissible control episodes are
country-year observations with the same recent treatment history that
remain non-democratic over the same post-treatment window. This design
mirrors the episode-based logic of Imai et al.
(\citeproc{ref-imai2023}{2023}) while expressing the final estimator in
weighted stacked DID form.

For this replication, only two estimators are used: CBWSDID and
PanelMatch, both of which rely on refinement. To make results more
comparable, the first-stage refinement is identical: nearest-neighbor
Mahalanobis matching with four control episodes and no replacement. The
matching variables follow the substantive covariate set used in the
Acemoglu et al.~as adapted in Imai et al.
(\citeproc{ref-imai2023}{2023}): four lags of the outcome, four lags of
log population, four lags of log population below age 16, four lags of
log population above age 64, four lags of net financial flows as a share
of GDP, four lags of trade volume as a share of GDP, and four lags of a
social-unrest indicator.

\begin{figure}

\centering{

\pandocbounded{\includegraphics[keepaspectratio]{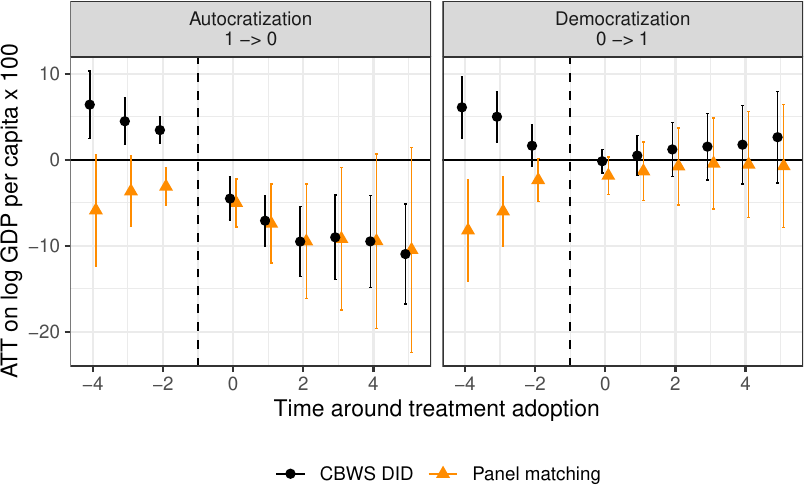}}

}

\caption{\label{fig-acemoglu-dynamic}Event-study comparison of CBWSDID
and PanelMatch in the democracy-and-growth application.}

\end{figure}

The results are presented in Figure~\ref{fig-acemoglu-dynamic}. It is
seen that for democratization episodes (\(0 \to 1\)), both estimators
imply weak growth effects. Specifically, the post-treatment coefficients
are close to zero in the short run and become only mildly positive at
longer horizons, with wide confidence intervals. In turn, for
autocratization episodes (\(1 \to 0\)), both estimators point to
persistently negative post-treatment effects and points estimates are
almost identical. It should be noted that the pre-treatment coefficients
remain nonzero, indicating that the design is demanding even with the
rich covariate set and rising concerns about comparability of treated
and control groups. Interestingly, both estimators, which converge in
post-treatment effects, display strikingly opposite pre-treatment
trends. The final point from this comparison is that the CBWSDID
estimator has lower variance estimates, though, one more time, the point
estimates are very similar.

\section{Conclusion}\label{conclusion}

This paper has proposed CBWSDID, a design-based extension of weighted
stacked DID for settings in which untreated trends may be conditionally
rather than unconditionally parallel. The main idea is simple but
useful: the problem of constructing a credible donor pool within each
sub-experiment should be separated from the problem of aggregating
treated and control observations across sub-experiments. Matching and
weighting address the first problem; corrective stacked weights address
the second. Expressing both stages through nonnegative design weights
makes it possible to combine them within a single regression-based
stacked estimator.

The same logic carries over to repeated-treatment settings once the
basic unit of analysis is redefined from first-adoption cohorts to
admissible treatment episodes. Under finite memory, the estimator can be
applied to \(0 \to 1\) and \(1 \to 0\) transitions without abandoning
the stacked-DID representation. This is important because many
substantive applications feature switching, reversals, and recurrent
exposure patterns that do not fit the standard absorbing-treatment
template.

The empirical evidence in the paper points to a consistent message. In
the simulation design, covariate balancing materially improves recovery
of the true dynamic effect path and sharply reduces spurious pre-trends.
In the Trounstine application, the contrast is even more striking:
unrefined estimators, including TWFE, Sun-Abraham, and weighted stacked
DID, all display sizeable pre-treatment drift, whereas both versions of
CBWSDID largely flatten the pre-period and substantially attenuate the
post-treatment pattern. In the Acemoglu democracy-and-growth
application, CBWSDID and PanelMatch deliver very similar substantive
conclusions, but CBWSDID does so within a weighted stacked DID framework
that remains easy to summarize, diagnose, and extend.

Taken together, these results suggest that CBWSDID is best understood
not as a replacement for modern DID estimators or panel matching, but as
a bridge between them. It preserves the transparent estimand and
aggregation logic of weighted stacked DID while importing the design
sensitivity of matching- and weighting-based refinement. For empirical
researchers working in demanding staggered or repeated-treatment
settings, that combination can be practically valuable.

\section*{References}\label{references}
\addcontentsline{toc}{section}{References}

\protect\phantomsection\label{refs}

\end{document}